\documentclass{aastex}
\usepackage{amssymb,amsmath, morefloats, natbib, graphicx,float,threeparttable,rotating,longtable,array}
\numberwithin{equation}{section}

\newcommand{\fig}[1]{{fig.\,\ref{#1}}}
\newcommand{\tab}[1]{{table\,\ref{#1}}}
\newcommand{\eqn}[1]{{eq.\,\ref{#1}}}

\newcommand{\eg}{{\it e.g.}}
\newcommand{\ie}{{\it i.e.}}

\newcommand{\todo}[2]{{}}

\newcommand{\pc}{\,\mathrm{pc}}
\newcommand{\au}{\,\mathrm{au}}
\newcommand{\km}{\,\mathrm{km}}

\newcommand{\meter}{\,\mathrm{m}}

\newcommand{\yr}{\,\mathrm{yr}}

\newcommand{\mags}{\,\mathrm{mag}}

\newcommand{\gps}{\ensuremath{g_{\rm P1}}}
\newcommand{\rps}{\ensuremath{r_{\rm P1}}}
\newcommand{\ips}{\ensuremath{i_{\rm P1}}}
\newcommand{\zps}{\ensuremath{z_{\rm P1}}}
\newcommand{\yps}{\ensuremath{y_{\rm P1}}}
\newcommand{\wps}{\ensuremath{w_{\rm P1}}}

\newcommand{\PSone}{\protect \hbox {Pan-STARRS1}}

\begin{document}

\title{An Observational Upper Limit on\\
the Interstellar Number Density of Asteroids and Comets}

\author{
Toni Engelhardt\altaffilmark{1,2},
Robert Jedicke\altaffilmark{1},
Peter Vere\v{s}\altaffilmark{1,3,4},
Alan Fitzsimmons\altaffilmark{1,5},\\
Larry Denneau\altaffilmark{1},
Ed Beshore\altaffilmark{6},
Bonnie Meinke\altaffilmark{1,7}
}


\altaffiltext{1}{Institute for Astronomy, University of Hawaii, Honolulu, HI, USA}

\altaffiltext{2}{Technical University of Munich, Munich, Germany}

\altaffiltext{3}{Comenius University in Bratislava, Bratislava, Slovakia}

\altaffiltext{4}{Jet Propulsion Laboratory, California Institute of Technology, 4800 Oak Grove Drive, Pasadena, CA 91109, USA}

\altaffiltext{5}{Queens University, Belfast, UK}

\altaffiltext{6}{The University of Arizona, Lunar and Planetary Laboratory,Tucson, AZ, USA}

\altaffiltext{7}{Space Telescope Science Institute, Baltimore, MD}

\section*{Abstract} 
\label{s.Abstract}

We derived 90\% confidence limits (CL) on the interstellar number density ($\rho_{IS}^{CL}$) of interstellar objects (ISO; comets and asteroids) as a function of the slope of their size-frequency distribution and limiting  absolute magnitude. To account for gravitational focusing, we first generated a quasi-realistic ISO population to $\sim750\au$ from the Sun and propagated it forward in time to generate a steady state population of ISOs with heliocentric distance $<50\au$.  We then simulated the detection of the synthetic ISOs using pointing data for each image and average detection efficiencies for each of three contemporary solar system surveys --- \PSone, the Mt. Lemmon Survey, and the Catalina Sky Survey.  These simulations allowed us to determine the surveys' combined ISO detection efficiency under several different but realistic modes of identifying ISOs in the survey data.  Some of the synthetic detected ISOs had eccentricities as small as $1.01$ --- in the range of the largest eccentricities of several known comets.  Our best CL of $\rho_{IS}^{CL} = 1.4 \times 10^{-4}\au^{-3}$ implies that the expectation that extra-solar systems form like our solar system, eject planetesimals in the same way, and then distribute them throughout the galaxy, is too simplistic, or that the SFD or behavior of ISOs as they pass through our solar system is far from expectations.

\section{Introduction} 
\label{s.Introduction} 

\todo{RJ}{REMEMBER TO SEND ARXIV LINK TO MIKE TURNER.}

Simulations of the formation and evolution of our solar system suggest that the early orbital migration of the gas and ice giant planets ejected up to 99\% of the original planetesimals into interstellar space \citep[\eg][]{Charnoz2003,Bottke2005}.  This scenario suggests that a large number of objects must occupy interstellar space yet a sizable, clearly interstellar object has never been identified \citep[on the other hand, tiny interstellar dust particles are well known \eg][]{Mann2010}. The spatial number density of interstellar objects (ISO) and their composition and size-frequency distribution would provide valuable information about the commonality of solar system formation processes such as the prevalence of giant Jupiter-like planets capable of ejecting planetesimals. In this work we calculate a limit on the ISO spatial number density using data from three contemporary wide-field solar system surveys.

We will use the term `interstellar object' (ISO) to mean asteroids and comets that are not gravitationally bound to a star. They almost always encounter our solar system on hyperbolic trajectories with heliocentric eccentricities significantly greater than 1.  Under exceptional circumstances, they may be captured through a gravitational encounter with Jupiter. \citet{Torbett1986} calculated that the ISO capture rate must be about 1 per 60 million years if the interstellar ISO number density\footnote{In this work we will always provide the ISO number density in interstellar space, far from the gravitational focusing of a stellar-mass body.  \citet{Torbett1986} neglected the gravitational deflection of ISOs by the Sun, and we will show that the ISO number density near Jupiter is only marginally higher than the interstellar value.} of objects with diameters $\ge 1\km$ is $10^{13}\pc^{-3}$ ($10^{-3}\au^{-3}$), which corresponds to each stellar system ejecting about $10^{14}$ `comets'.  For perspective, that interstellar ISO spatial density corresponds to about 2 ISOs in a sphere with the diameter of Saturn's orbit without accounting for the Sun's gravitational focusing.  Given that the average dynamical lifetime of short period comets is about 0.45 million years \citep{Levison1994}, the steady state number\footnote{The steady state number of objects in a population ($N$) is related to the flux ($F$) of objects entering or leaving the population and their mean lifetime ($L$) as members of the population by the $N=FL$ equation.} of captured ISOs in the solar system should be about 0.01, or, roughly a 1\% chance of there being a captured ISO at any time.  The corollary to this statement is that most ISOs that are present in the solar system are unbound.

Comet 96P/Machholz is  currently the best candidate for being an interstellar interloper \citep{Schleicher2008} because 1) it is the only known short-periodic comet with both high orbital inclination and high eccentricity and 2) it has an unusual composition, being depleted in both carbon and cyanogen, which suggests a different origin from other known comets.  Backward propagation of 96P/Machholz's orbit can not establish it as an ISO due to close approaches with giant planets that cause chaotic jumps in semi-major axis which, combined with a Kozai resonance with Jupiter, may lead to a Sun impact within the next $1.2\times10^4$ years \citep{Gonczi1992,FuenteMarcos2015}.

One of the problems in comparing previous theoretical and observational estimates of the ISO number density is that there is no agreement on the size range at which the value should be calculated or quoted.  We suggest that ISOs include all objects regardless of whether they are asteroids or comets and that the number density refers to those larger than $1\km$ diameter, corresponding roughly to the size of a typical comet \citep{Weissman1983}.  We have attempted to convert earlier estimates to our canonical suggestion using a cumulative size-frequency distribution (SFD) of the form $N(<H) \equiv N(H) \propto 10^{\alpha H}$ or $N(>D) \equiv N(D) \propto D^{-a}$ where $\alpha=0.5$ and $a=2.5$ are the slopes of the distributions for a self-similar collisional cascade \citep[\eg][]{Dohnanyi1969,Durda1993}, and $H$ and $D$ represent the objects' absolute magnitudes and diameters respectively.  We note that the Jupiter family comet (JFC) SFD has $a\sim1.9$ \citep[or, equivalently, $\alpha\sim0.38$; \eg][]{Snodgrass2011, Fernandez2013} but, as the ISO SFD is completely unconstrained, the use of the theoretical value is justified.  We also note that JFC nuclei typically have measured  $D>1\km$, but these studies are significantly biased against observation of small nuclei.

The ISO number density in interstellar space ($\rho_{IS}$) can be estimated by 1) assuming that our solar system is typical, 2) using numerical simulations to calculate the number of objects that were ejected from our solar system, 3) multiplying that number by the number of star systems in the galaxy, 4) assuming that the galactic orbits of the ejected objects are randomized through galactic tides and stellar encounters, and 5) dividing by the volume of the galaxy. Roughly this technique yields $\rho_{IS} \sim 10^{-3}\au^{-3}$ \citep{McGlynn1989,Jewitt2003} while \citet{Sen1993}'s more detailed estimate predicts about a sixth of that value with $\rho_{IS} \sim 1.6 \times 10^{-4}\au^{-3}$. These values are considerably larger than the $5\times10^{-9}$ to $5\times10^{-5}\au^{-3}$ range predicted by \citet{Moro-Martin2009}\footnote{\citet{Moro-Martin2009} provided their estimates for the number of objects with {\it radius} $>1\km$ and we have corrected their values to our $1\km$ {\it diameter} standard by including a factor of $5\sim2^{2.5}$, roughly correcting for the size-frequency distribution in the $1\km$ diameter range according to the SFD expected for a self-similar collisional cascade \citep{Dohnanyi1969}. We note that \citet{Moro-Martin2009} implemented several different SFD slopes depending on the object's type and size and also assumed an albedo of 6\% compared to the 4\% used here.} who suggest that the earlier values are too high due to neglecting important factors such as stellar mass and the presence of giant planets in the star system.

Some experimental measurements of the ISO number density have relied on indirect techniques. \citet{Jura2011} calculated the number density of ISOs using the hydrogen content in helium-dominated atmospheres of hydrogen-depleted white dwarfs.  Their analysis assumes that the present hydrogen is delivered by ISOs rather than an {\it in situ} debris disk and they claim that their results exclude `optimistic' ISO number densities but can not exclude the \citet{Moro-Martin2009} estimate.  Another study suggests that Sgr~A* flares could be induced by asteroids or comets with radii larger than $10\km$ \citep{Zubovas2012}. 

Not a single macroscopic object has ever been established as an ISO despite the fact that dedicated surveys of the solar system have been operating for almost three decades \citep[\eg][]{Jedicke2015AIV}.  They were established to discover and to monitor large near-Earth objects (NEO) but have been very successful at detecting comets and all classes of asteroids from the main belt to the trans-Neptunian region.  The current generation of surveys have fainter limiting magnitudes and are capable of surveying a significant fraction of the sky on a nightly basis, and the next generation will provide even deeper images over wider areas (\eg\ LSST, \citet{Ivezic2008}; SST, \citet{Monet2013}).  The larger search volume of the new surveys will provide a slightly better chance of detecting ISOs compared to existing surveys \citep[\eg][]{Cook2011}.


Some marginally hyperbolic objects have been detected:  the JPL Small-Body Database Search Engine lists\footnote{as of 2016 October 14} 292 objects with $e>1.0001$, five objects with $e \ge 1.01$, and the highest eccentricity object is C/1980~E1~(Bowell) with $e=1.0577$.  Conventional wisdom suggests that their barely hyperbolic orbits originate either in perturbations by the solar system's planets during their passage through the solar system \citep[\eg][]{Buffoni1982} or perhaps through the effect of out-gassing from the nucleus.  However, we will show below that these comets' eccentricities are within the range of possible ISO eccentricities.

\citet{Afanasiev2007} made the astounding claim of detecting and obtaining a spectrum of a centimeter-scale {\it intergalactic} meteor using a multi-slit spectrometer on the $6\meter$ Special Astrophysical Observatory of the Russian Academy of Sciences.  They further claim that observations with a wide field camera identified a dozen meteors consistent with the expected radiant for intergalactic objects coming from the direction of motion of the Milky Way through the Local Group of galaxies.  Their suggestion that about 5\% of the meteors they detected were intergalactic in origin is inconsistent with the lack of any supporting evidence from other optical and meteor radar observatories \citep[\eg][]{Musci2012,Weryk2004}.

\citet{Francis2005} utilized the long-periodic comet population's detectability with the LINEAR survey \citep{Stokes2000} to derive a 95\% upper confidence limit on the ISO number density of $\rho_{IS}^{CL} \sim 4.5 \times 10^{-4}\au^{-3}$.  In this work, we improve and extend the technique using a synthetic ISO population and modeling the combined ISO detection efficiency for three long-term contemporary surveys, the Catalina Sky Survey and Mt. Lemmon Survey \citep{Christensen2012}, and \PSone\ \citep{Kaiser2010}.

\section{Survey data} 
\label{s.SurveyData} 
 
\subsection{\PSone} 
\label{ss.PS1}
The \PSone\ telescope \citep[MPC Code F51;][]{Kaiser2010} on Haleakala, HI, is a prototype of a next generation all-sky survey telescope \citep{Kaiser2002} designed to explore the observable universe from interior to Earth's orbit out to cosmological distances.  The $1.8\meter$ $f$/4 Ritchey-Chretien optical assembly and 1.4\,gigapixel camera \citep{Tonry2009} provide a $\sim7\deg^2$ field-of-view at $\sim0.26\arcsec$/pixel. The camera consists of an array of 60 CCDs that each consist of an $8\times8$ array of $600\times600$ pixel `cells' that can be read in parallel. The system now devotes 90\% of its time to surveying for NEOs but in early 2014 it completed a 3-year survey of the sky north of $\sim-30\arcdeg$ declination in 5 Sloan-like filters \citep{Fukugita1996}.  The primary filters (\gps, \rps, \ips, \zps\ and \yps)  cover the visible to NIR spectrum \citep{Tonry2012b,Schlafly2012,Magnier2013} and the wide-band filter ($\wps\sim\gps+\rps+\ips$) was specifically designed to maximize the NEO detection efficiency.  

The 3-year \PSone\ survey had 5 distinct components, but most of the data were suitable for detecting asteroids and comets. The main $3\pi$-steradian survey mode \citep{Schlafly2012} required $\sim56$\% of the survey time in the 5 primary filters. In this mode, the same field was visited $2\times$ or $4\times$ within a night in 30 to $40\sec$ exposures with a total time separation of about an hour. The time between two visits to the same field, a transient time interval (TTI), was typically $15\min$. The medium deep survey (MD), with 25\% of the survey time focused on 10 fields of cosmological and extragalactic interest \citep{Tonry2012a}, was also suitable for identifying solar system objects.  The MD survey visited a single field $8\times$/night with filter-dependent exposure times of 120 to $240\sec$ in the $3\pi$ filters. The solar system survey (SS) used 5-6\% of the survey time but was increased to 12\% of the survey time after 2012 \citep{Denneau2013}.  It used the \wps\ filter with $45\sec$ exposures and mostly visited fields near opposition, or the `sweetspots' near the ecliptic at solar elongations of $60\arcdeg$ to $90\arcdeg$. In the SS survey each field was visited $4\times$ with $\sim20\min$ TTI near opposition and $\sim7\min$ TTI in the sweetspots. The SS survey was the most successful one in terms of limiting magnitude and solar system object discoveries and detections. 

During its 3-year survey, \PSone\ discovered more than 800 NEOs, $>40,000$ other asteroids, almost 50 comets, reported $\sim7,200,000$ asteroid positions, and observed $\sim560,000$ distinct asteroids \citep[\eg][]{Wainscoat2013,Veres2015}.  This study used \PSone\ observations between February 2011 and June 2013.

\subsection{Catalina and Mt. Lemmon Surveys} 
\label{ss.CSS}
The Catalina Sky Survey consists of the $0.7\meter$ Schmidt telescope, hereinafter referred to as CSS (MPC Code 703), and the $1.5\meter$ Mt. Lemmon reflector, or MLS, (MPC Code G96). Located north of Tucson, Arizona, both survey nightly for NEOs, except for approximately 2 months during the southwest's summer monsoon season and for about 6 to 7 days centered on full moon. Both telescopes employ automated image analysis and moving object detection software pipelines followed by same-night manual review of all detections.  This process  allows for same-night followup thereby helping to ensure that fast-moving or faint objects can be recovered on subsequent nights. The wide-field CSS covers $\sim8\deg^2$ with each image, allowing it to observe most of the night sky during a single lunation. The deeper but narrower field MLS survey with its $\sim1\deg^2$ field of view concentrates its observations near opposition or along the ecliptic. The images from both telescopes are un-filtered to maximize throughput and discovery statistics. In 2014, CSS and MLS accounted for just over 41 percent of all new NEO discoveries (\url{http://neo.jpl.nasa.gov/stats/}). This study used all fields acquired by these surveys until the end of 2012, beginning in Feb 2005 for MLS and in Jan 2005 for CSS.

\section{Synthetic ISO population} 
\label{s.SyntheticISOpopulation}

Our goal is to set an observational upper limit on the steady-state, interstellar, spatial number density of ISOs using the fact that \PSone, MLS, and CSS did not detect a single ISO in about 19 cumulative survey-years. To do so requires determining the combined ISO detection efficiency of the three surveys. We accomplished this measurement using a synthetic ISO population that was run through a survey simulation using actual fields observed by \PSone, MLS, and CSS, and each survey's average detection efficiency as a function of apparent magnitude for the appropriate filters.

\subsection{ISO orbit distribution}
\label{ss.ISOorbitDistribution}

Our ISO model expands upon the technique developed by \citet{Grav2011} that includes the propagation and gravitational focusing through our solar system of an originally homogeneous and random population of synthetic ISOs in a large heliocentric sphere with radius $r_0$. We generated random positions for the synthetic ISOs within the sphere at $t_0$ (we will use $t$ to indicate a specific time and $T$ to represent a time duration) and assigned them random direction vectors with random, Gaussian-distributed speeds.  We refer to this population as the synthetic `generated' population. The relative speed of ISOs with respect to the Sun is expected to be of the same order as that of nearby stars with a mean speed of $\bar v = 25\km\sec^{-1}$ and $\sigma = 5\km\sec^{-1}$ \citep[\eg][]{Grav2011,Kresak1992,Dehnen1998}.  This distribution implies that 99.7\% of the ISOs have interstellar speeds relative to the Sun between $v_{min} = 10\km\sec^{-1}$ and $v_{max} = 40\km\sec^{-1}$.

The spatial and velocity distributions of the synthetic ISOs within the sphere with radius $r_0$ at time $t_0$ (both parameters to-be-determined below) are probably a fine representation of their steady-state distributions in interstellar space but are not at all representative of their steady-state spatial and velocity distribution in the inner solar system due to gravitational focussing by the Sun.  We generated a steady-state ISO population, `the model', within a `core' sphere of radius $r_{model} = 50\au \ll r_0$ centered on the Sun, by propagating the trajectories of the synthetic generated interstellar ISO population forward in time. The $50\au$ value was chosen because an ISO would have to be several hundred kilometers in diameter to be detected by any of the three surveys and an ISO of this size within that distance is extremely improbable.  To ensure that our steady-state model within the solar system is representative of the expected distribution, the slowest synthetic objects at $t_0$ within $r_{model}$ must be able to exit the core, and the generated volume must be larger than the distance that can be traveled by the fastest objects in the model. We propagated the interstellar model for a `preparation time'
\begin{equation}
T_{prep} \ge { 2 \, \frac{r_{model}}{v_{min}} } 
\end{equation}
and ensured that 
\begin{equation}
r_0 \ge v_{max} \; T_{prep} 
\end{equation}
where the latter formula intentionally ignores the ISOs' acceleration due to the Sun and assumes that they are on a direct path to the heliocenter.  We used $T_{prep} \sim 70\yr$ and $r_0 \sim 750\au$ that both include margins of about 50\%. 

The ISO model must represent the steady-state distribution of ISOs in the inner solar system during the combined survey time range of the three surveys, $T_{survey}=t_f-t_i$, where $t_i=53371$\,MJD and $t_f=57387$\,MJD corresponding to the time period from 2005 January 1 through 2015 January 1 that brackets the actual surveys' duration.  Thus, $t_0 = t_i - T_{prep}$ which we set to 27399\,MJD (1933 Nov 23).

We generated about 1.7\,billion synthetic ISOs within $r_0$ at $t_0$ and preselected those that would be in the model (core) volume during the survey time as calculated using the hyperbolic Kepler equation.  We also eliminated the relatively small number of non-hyperbolic synthetic objects with $e<1$ at $t_0$ that are artifacts of the synthetic ISO generation technique.  Finally, we propagated the $\sim 1$ million remaining objects to the surveys' starting time, $t_i$, with the OpenOrb n-body integrator \citep{Granvik2009} incorporating the Sun, major planets, Pluto, and the Moon, using the DE405 planetary ephemerides \citep{Standish1998}. About 35\% of the $\sim$1,000,000
synthetic model ISOs are located in the model $50\au$ sphere at any time so their average number density is slightly higher than the simulation's interstellar value of about $0.66\au^{-3}$.

\begin{figure}[ht!]  		
\centering	
\includegraphics[width=\textwidth]{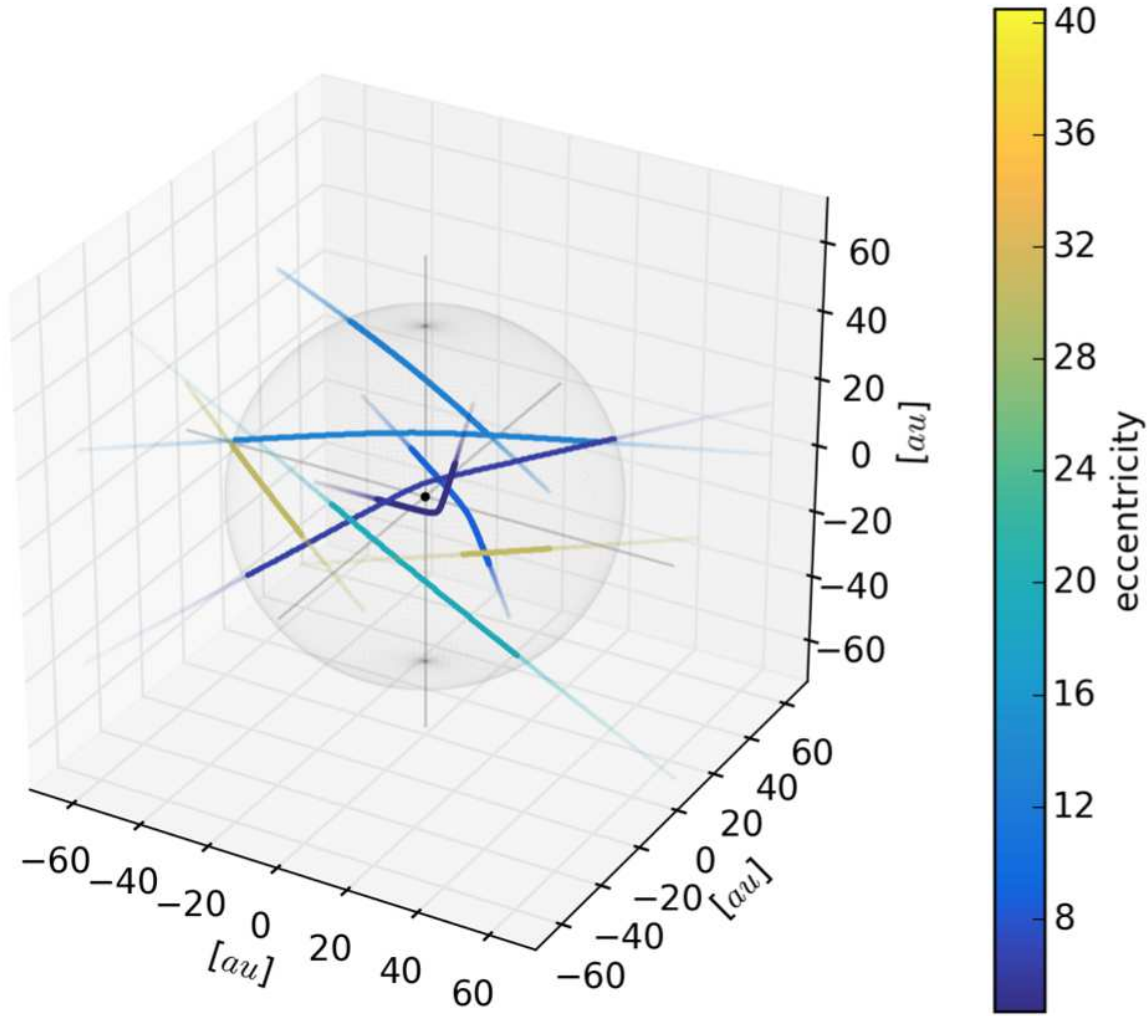}
\caption{Trajectories of 8 synthetic ISOs within the $50\au$ radius model centered on the Sun with 100 day sampling. The two objects with the smallest eccentricities, with the smallest heliocentric distance and the most curvature in their trajectories, have $e\sim4$ and $e\sim6$ while the other six objects were selected to represent a wide range of ISO trajectories.
}	
\label{fig.8-synthetic-ISO-trajectories}
\end{figure}

\begin{figure}[!ht]
\centering
\includegraphics[width=.45\textwidth]{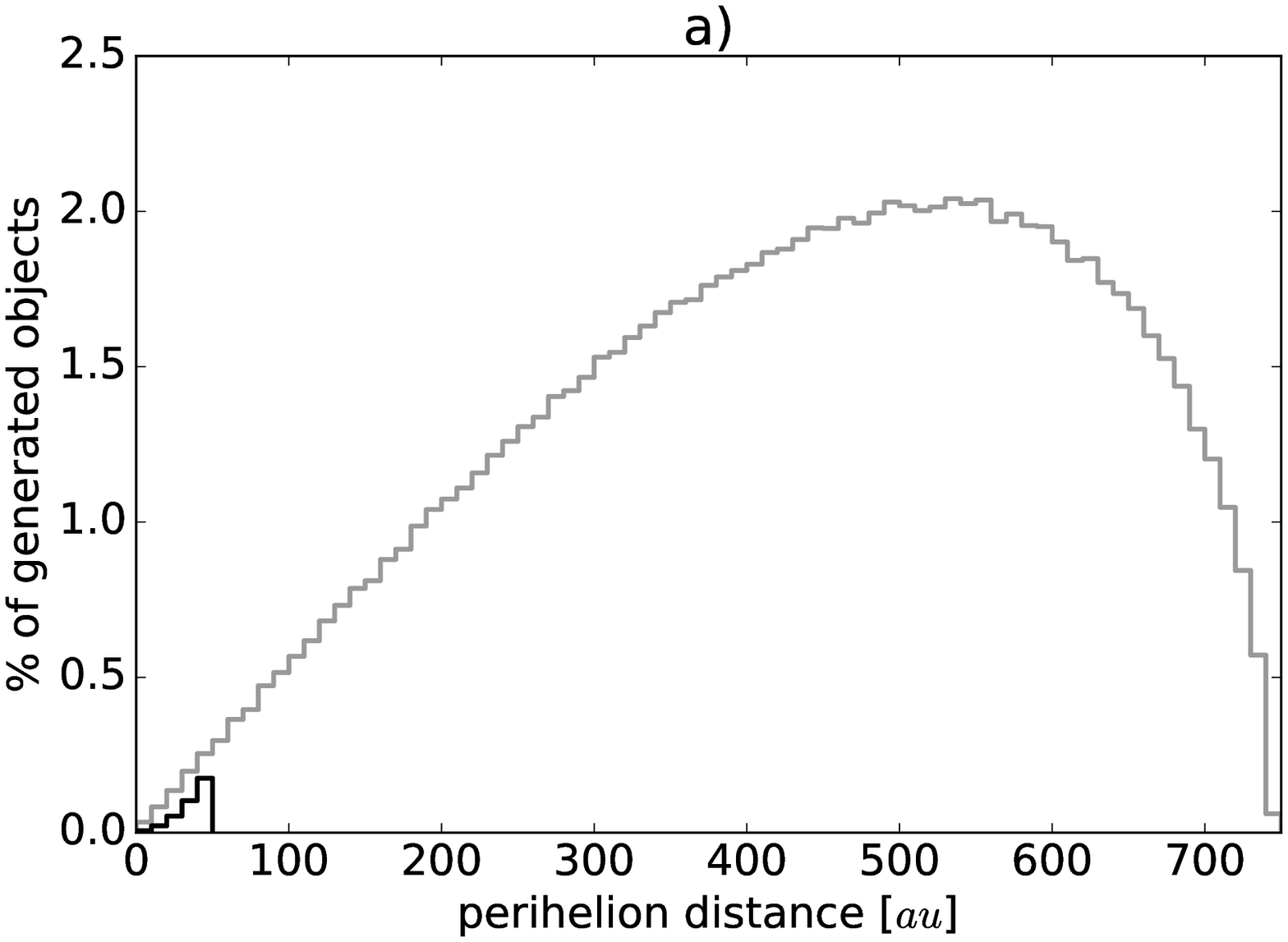}
\includegraphics[width=.45\textwidth]{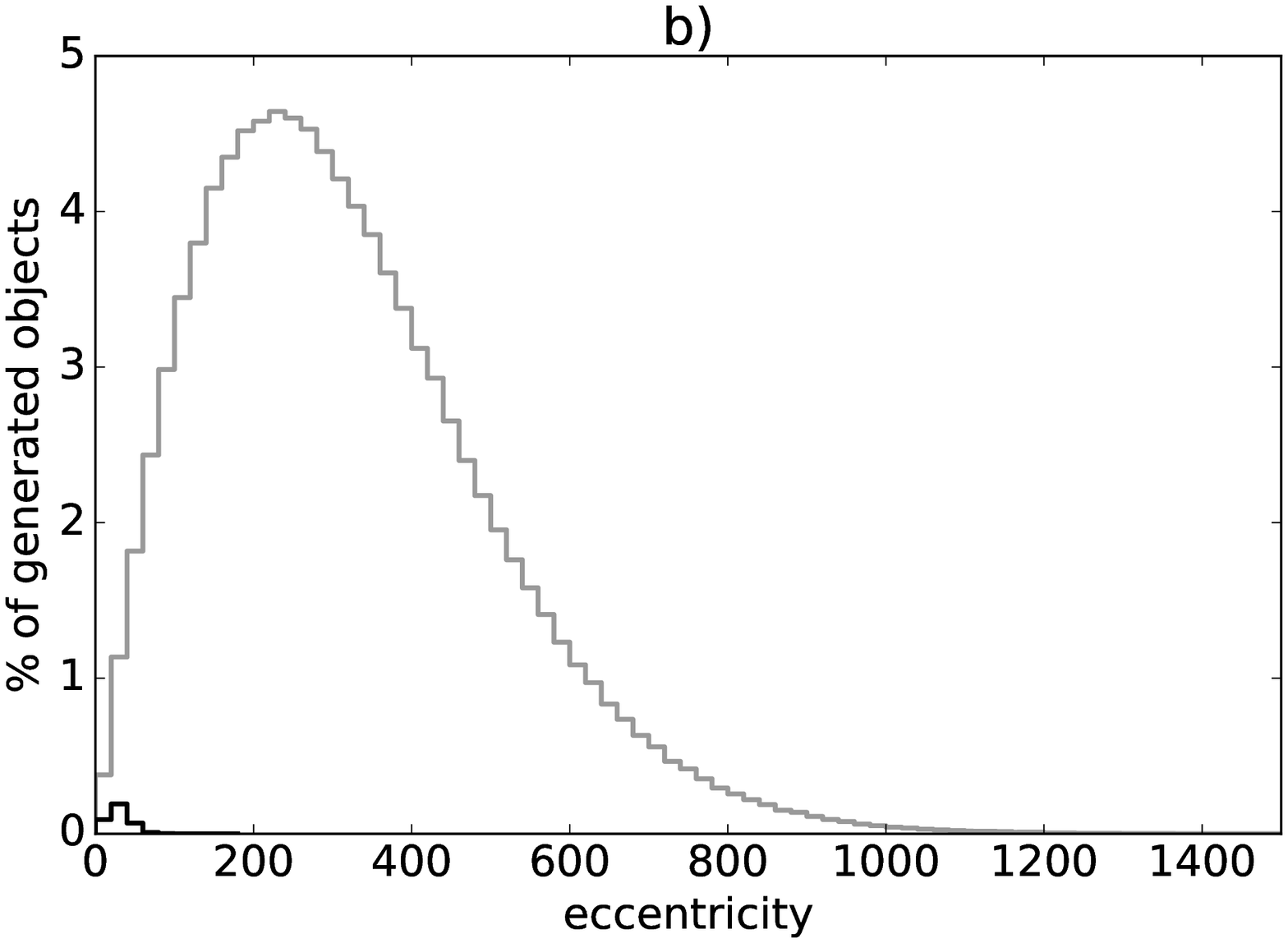}
\includegraphics[width=.45\textwidth]{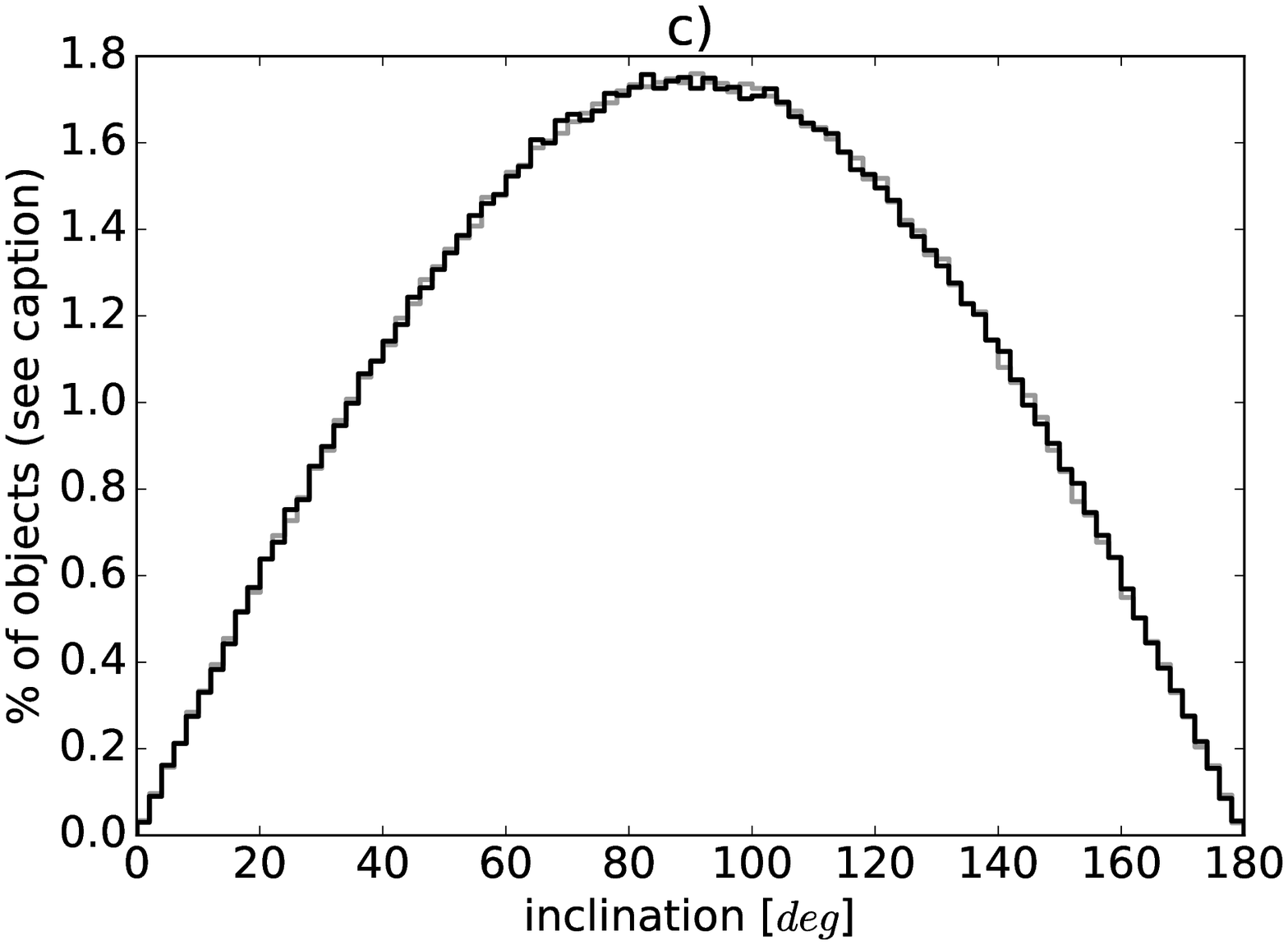}
\includegraphics[width=.45\textwidth]{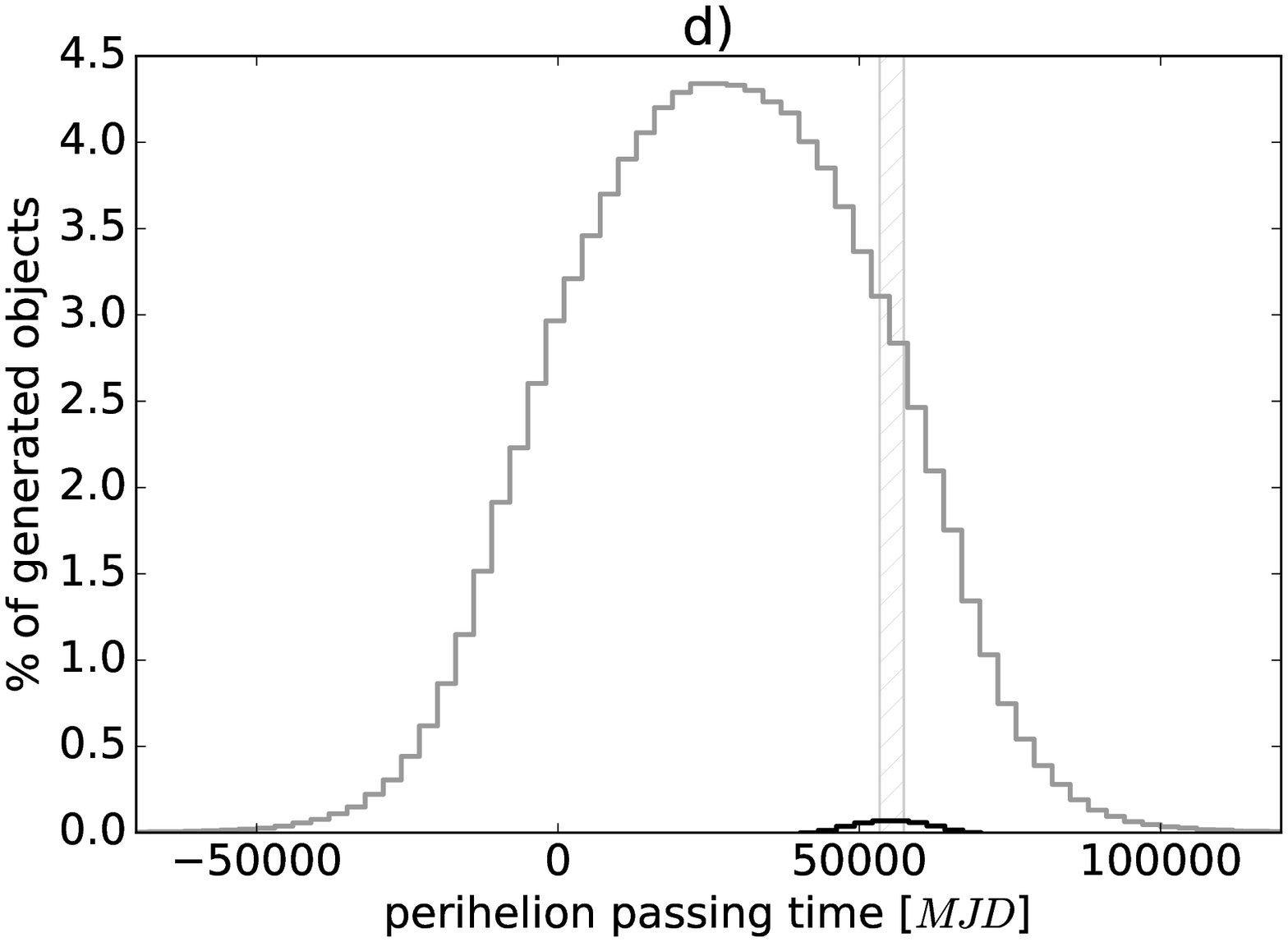}
\caption{Panels a, b, and d) The generated (gray) and model (black) ISO orbital parameter distributions as a percentage of the generated distribution as a function of a) perihelion distance, b) eccentricity, and d) time of perihelion.  In panel c) the ISO model (black) inclination distribution is provided as a percentage of the model itself.}
\label{fig.ISO-orbit-distribution}
\end{figure}

An ISO's eccentricity is related to its perihelion distance and speed (\fig{fig.8-synthetic-ISO-trajectories} and \fig{fig.ISO-orbit-distribution} a \& b). The larger its perihelion or the faster it moves relative to the Sun, the less its trajectory is modified by gravitational acceleration, and the higher its eccentricity. Thus, very distant ISOs will follow nearly straight lines and have eccentricities approaching infinity. Conversely, the closer an ISO approaches the Sun and the slower it moves, the lower the eccentricity. The perihelion distance of generated objects peaks at about $500\au$ because of the truncation at $r_0\sim750\au$ \ie\ it is unlikely to randomly generate objects with perihelia just inside the maximum distance (\fig{fig.ISO-orbit-distribution}a).  The model object population has a smooth increase in the distribution of perihelion distances as expected.  The eccentricity distribution of the generated objects peaks at about 250 but has a long tail extending to $e>$1,000, while the model ISOs, those that enter the $50\au$ radius heliocentric sphere within the survey time, have a maximum at $e\sim25$ and only about 0.1\% have $e<2$.  However, ISOs with $e\ga1.0$ are exactly the ones with small perihelion that are most likely to be detected by surveys, and in our simulation about 0.003\% of the model ISOs had $e \le 1.0577$.  Thus, there is a small probability that ISOs which pass close to the Sun may appear to be on gravitationally perturbed Oort cloud like orbits.

The potentially observable model ISOs have essentially the same inclination distribution as the generated population because the selection criteria are independent of the inclination. Even if we had selected the observable model ISOs with a full n-body propagation into the solar system the inclination distributions would have been nearly indistinguishable because only extremely rare encounters with planets would affect the objects' inclinations (\fig{fig.ISO-orbit-distribution}c).  The distribution is a simple $\sin i$ function due to the phase space of available normals to the orbital planes as a function of inclination.  The time of perihelion ($t_p$) for the generated population is pseudo-normally distributed around $t_0=27399$\,MJD by design as described above (\fig{fig.ISO-orbit-distribution}d).  Similarly, $t_p$ for the model population is distributed pseudo-normally around the time period during which the survey data used in this work was acquired. The model was designed such that within the survey duration the distribution of the times of perihelion passages is essentially flat, as would be expected for objects making one passage through the solar system.

\begin{figure}[!ht]
\centering
\includegraphics[width=0.8\textwidth]{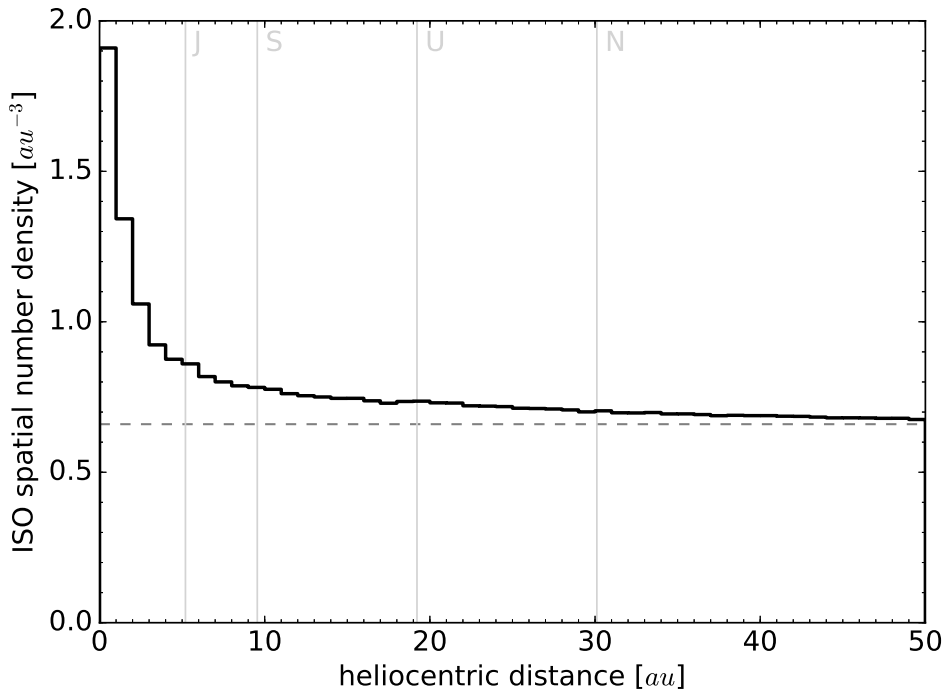}
\caption{
Average ISO incremental number density in shells of $1\au$ thickness versus heliocentric distance during the survey period.  Note the asymptotic approach to the model's interstellar number density of about $0.66\au^{-3}$ indicated by the horizontal dashed grey line.  Vertical gray lines represent the semi-major axes of Jupiter (J), Saturn (S), Uranus (U) and Neptune (N).  
}
\label{fig.ISOnumberDensity.vs.heliocentricDistance}
\end{figure}

The spatial number density of the model synthetic ISOs decreases asymptotically with heliocentric distance (\fig{fig.ISOnumberDensity.vs.heliocentricDistance}) and is essentially equal to the interstellar value at $50\au$ \ie\ at TNO-like distances.  The density increases near the Sun due to gravitational focusing and is about $3\times$ higher than the interstellar value within about $1\au$ of the Sun \ie\ within Earth's orbit.  Note that the ISO spatial density at Jupiter's distance from the Sun, about $5.2\au$, is about 30\% higher than the interstellar average.

\section{The Moving Object Processing System (MOPS) \\
and ISO discovery efficiency} 
\label{s.MOPS+ISOdetectionEfficiency}

The \PSone\ MOPS can link multiple observations of the same object together within a night into `tracklets', combine tracklets from different nights into `tracks', calculate orbital elements, perform attribution of new tracklets to known objects, identify `precoveries' of historical tracklets associated with newly calculated orbits, and allow for manual vetting of all the data \citep{Denneau2013}.  We used MOPS to simulate the detection of our synthetic ISOs in \PSone, CSS, and MLS fields.  In particular, we used MOPS to determine which of the million synthetic ISOs appeared in each of the 181,388 \PSone\ fields, 244,854 CSS fields and 208,464 MLS fields, as well as determining their heliocentric and geocentric distance at the time of each observation and their `interesting object score' (described later in this section).  We use these values to calculate the probability that the object will be identified as an ISO candidate once we assign the ISO a diameter (or absolute magnitude).  Tracklets for the model ISOs that have non-zero detection efficiency comprise the set of `detectable' objects.

Each system's time-averaged tracklet detection efficiency was fit to the empirical function
\begin{equation} 
\label{eq.TrackletEfficiency}
\epsilon_{F}(m_F) =
\left\{
  \begin{array}{c l}
    \epsilon_{0F} \Big[ 1+e^{\frac{m_F-L_F}{w_F}} \Big]^{-1} & \mbox{if } m_F \le L_F \\
    0 & \mbox{otherwise}
  \end{array}
\right.
\end{equation}
where $m$ is the objects' apparent magnitude, $\epsilon_{0}$ is the maximum detection efficiency for bright objects, $L$ is the apparent magnitude at which the efficiency drops to 50\% of its maximum {\it and} the limiting apparent magnitude at which we set the detection efficiency to zero, $w$ is a measure of the range of apparent magnitudes over which the efficiency drop occurs, 
and the $F$ sub-scripts indicate that each parameter is filter dependent (see \tab{tab.SurveyEfficiencyParameters}).  
We impose $\epsilon_{0F}=0$ for $m_F > L_F$ because without this requirement \eqn{eq.TrackletEfficiency} allows for small efficiencies at faint apparent magnitudes where the size-frequency distribution would predict a large number of objects and this scenario can allow unrealistically faint objects to be detected in the simulation. 
\begin{table}[htp]
\centering
\begin{tabular}{|c|c|c|c|c|}
\hline
Survey (obs. code)    &  Filter  &  $\epsilon_{0F}$  &  $L_F$  &  $w_F$ \\
\hline  
          &   g      &  0.69             &  20.1   &  0.22  \\
          &   i      &  0.66             &  20.5   &  0.24  \\
PS1 (F51) &   r      &  0.67             &  20.5   &  0.23  \\
          &   w      &  0.68             &  21.3   &  0.27  \\ 
          &   y      &  0.53             &  18.7   &  0.21  \\
          &   z      &  0.55             &  19.8   &  0.20  \\
\hline
CSS (703) &   none   &  0.70             &  19.4   &  0.39  \\
\hline
MLS (G96) &   none   &  0.85             &  21.1   &  0.42  \\
\hline
\end{tabular}
\caption{Filter and survey dependent efficiency parameters (see \eqn{eq.TrackletEfficiency})}
\label{tab.SurveyEfficiencyParameters}
\end{table}

There are numerous caveats that could be discussed in regard to using or calculating the surveys' tracklet detection efficiency.  In particular, the range of apparent rates of motion over which the quoted efficiency (\eqn{eq.TrackletEfficiency}) is valid is mostly restricted to values typical of main belt asteroids simply because those are the most numerous objects from which the efficiency is measured.  The fact that the surveys regularly identify objects moving at both much faster (NEO) and slower (Centaur) rates suggests that it is not inappropriate for us to apply the efficiency function over a wider range of rates of motion.  But our laissez-faire application clearly has its limits at both fast and small rates of motion and there is also a secondary-dependence on the seeing.  For instance, \PSone\ and MOPS detect transient objects through subtraction of consecutive images\footnote{The CSS and MLS surveys do not employ image differencing and are not affected by this limitation.}. If an object moves less than about a seeing disc between images they will be `self-subtracted' with a concomitant reduction in detection efficiency.  However, any ISO is likely to remain visible for months or years, it is unlikely that {\em no} revisits to the object will be in good observing conditions, and nights of better seeing naturally correspond to fainter limiting magnitudes and better sensitivity. Thus, we consider our efficiency parameterization sufficient for setting a limit on the population for the `typical' ISOs that might actually be detectable with one of the three surveys.

The discovery of an ISO requires not only that the tracklet be identified in a set of images but also that it be recognized as an interesting candidate worthy of followup and confirmation as an ISO.  The typical technique for identifying a tracklet as an unknown NEO candidate by the three NEO surveys is to use the MPC `digest' score ($p$), a pseudo-probability that a tracklet is `interesting'.  It depends upon an object's apparent angular speed $\omega$, apparent magnitude, and apparent position relative to opposition.  The visible ISOs typically have high digest scores by virtue of their isotropic inclination distribution and high speeds as they pass through our solar system. But if they do not have high digest score, perhaps because they happen to appear with main belt like rates of motion in or near the ecliptic, it is unlikely that they would be detected as interstellar.  In practice, tracklets submitted to the MPC with $p\ga0.9$ receive enough followup effort to effectively guarantee that an ISO would be identified if it were hidden amongst NEO and other interesting candidates.

\begin{figure}[!ht]
\centering
\includegraphics[width=0.8\textwidth]{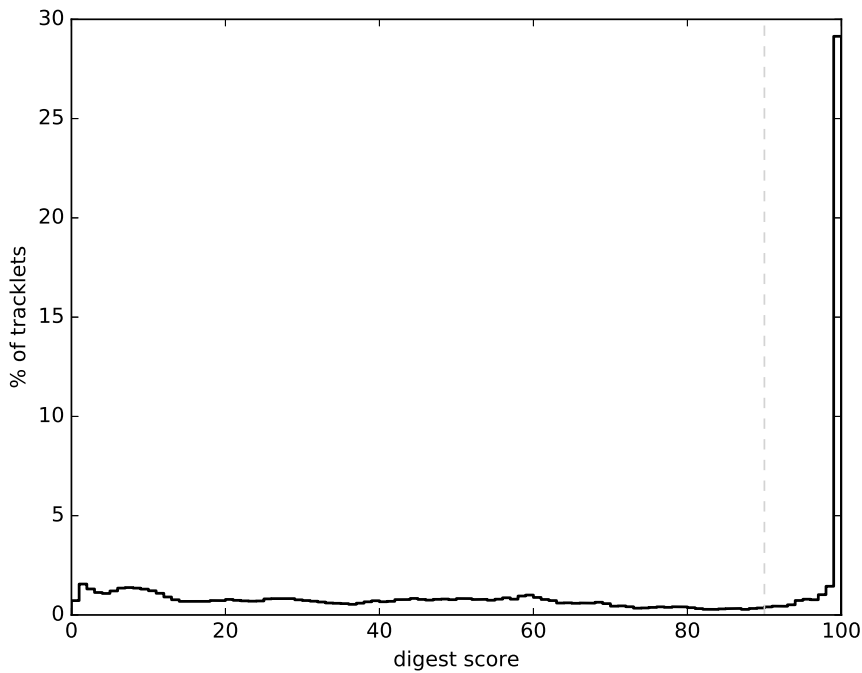}
\caption{The Minor Planet Center's `digest' score for detectable synthetic `cometary' ISO tracklets in our simulation.  The dashed line at digest=90 represents the limiting value above which a tracklet becomes `interesting' enough to trigger followup observations.}
\label{fig.ISO-DigestScores}
\end{figure}

We were surprised to find that roughly 2/3 of {\it detectable} `cometary' ISO tracklets have digest scores of $<90$ (\fig{fig.ISO-DigestScores}) making them `uninteresting' and unlikely to be targeted for followup by the many professional and amateur astronomers around the world who regularly provide this service \citep{Jedicke2015AIV}. The nearly flat distribution of digest scores $<90$ suggests that the majority of ISO tracklets usually have only mildly interesting apparent rates of motion or locations on the sky.  However, the same object could appear in different fields and different tracklets for the same object can have different digest scores.  In this case, it could be argued that the most important tracklet for ISO discovery are only those with the highest digest score for each object.  In this case, about 2/3 of the detectable ISOs had a maximum digest score $>90$.  This fact suggests that future sky surveys like LSST \citep[\eg][]{Ivezic2008} that will provide self-followup of their own discoveries could detect ISOs 50\% more efficiently than contemporary surveys because they will employ automated tracklet linking and orbit determination.  Finally, the `cometary' ISOs may be identified by their morphology on the image (presence of coma or tails) so that digest scores provides a lower limit to the detectability of active ISOs (we discuss this possibility below and use it to set our most stringent upper confidence limit on the interstellar ISO number density).  

The combined probability of discovering the $j^{th}$ ISO in the synthetic model by the three surveys is then
\begin{equation} 
\label{eqn.ISOdiscoveryEfficiency}
	\epsilon_j 
	  = 1 - 
	    \prod_F \; 
	      \bigl[\; 
	        1 - \epsilon_{jF} \; \mathcal{H}(p_j-0.9)
	      \bigr] 
\end{equation}
where $\epsilon_{jF}$ is the system's tracklet detection efficiency for that ISO in filter $F$ (\eqn{eq.TrackletEfficiency}), $p_j$ is the digest score for that tracklet, and we have introduced the Heaviside function with $\mathcal{H}(x)=0$ for $x<0$ and $\mathcal{H}(x)=1$ for $x\ge0$.  

The total number of synthetic ISOs detected in the simulation is simply
\begin{equation} 
\label{eqn.N_ISO}
	N^{*} = \sum_j \; \epsilon_j
\end{equation}
where we use the asterisk to denote the synthetic population and simulation. We can generalize the expression to the total number of synthetic detected ISOs with maximum absolute magnitude $<H_{max}$ when they have a size-frequency distribution $\propto 10^{\alpha H}$:
\begin{equation} 
\label{eqn.N_ISO.vs.alpha+H}
	N^{*}(\alpha,H_{max}) = \sum_j \; \epsilon_j(\alpha,H_{max}).
\end{equation}

\subsection{ISO apparent magnitudes}
\label{ss.ISO-V}

The ISO orbit distribution described above (\S\ref{ss.ISOorbitDistribution}) is intentionally size-independent so that we can assign any size or absolute magnitude to any object in the model.  So we assigned each object the same and arbitrary absolute magnitude, $H_0=0$, far larger than any object that we could expect to detect with the surveys in the limited survey duration, to determine which fields the ISOs would appear in if they were bright enough, and their apparent magnitude $V_0$, geocentric distance $\Delta$, heliocentric distance $r$, and rate of motion $\omega$ at the time of each observation.  MOPS calculates the asteroidal apparent magnitude according to \citet{Bowell1988}:
\begin{equation} 
\label{eq.Bowell}
  V_0 = H_0 
      + 5 \log (r\Delta) - 2.5 \log \big[ (1-G)\,\Phi_1 + G\,\Phi_2 \big]
\end{equation}
where the slope is fixed at the standard value of $G=0.15$, and $\Phi_1$ and $\Phi_2$ are phase functions that depend on an object's phase angle.    These parameters are then used to determine a synthetic ISO's digest score and apparent magnitude ($V$) when assigned any other absolute magnitude ($H$): $V=V_0+H$.


It is likely that most ISOs will contain some volatile material and display cometary activity which would cause them to be brighter than predicted using the standard asteroidal formula (\eqn{eq.Bowell}).  The diameter of an inert (asteroidal) ISO with a cometary geometric albedo of $p_V=0.04$ is given by
\begin{equation}
\frac{D}{\km} = 665\times 10^{-H/5} \; p_V^{-1/2} 
              \equiv 3325\times 10^{-H/5}
\label{eq.Diameter.vs.albedo+H}
\end{equation}
Under the assumption of either slow rotation and/or negligible thermal inertia, ignoring heat conduction into the interior, and a heliocentric distance of $1\au$, simple sublimation theory \citep{Cowan1979} predicts that the sunlit hemisphere of a water-ice dominated cometary nucleus will emit $4.4\times10^{28}$ H$_2$O molecules sec$^{-1} \meter^{-2}$.  The total sublimation rate will scale with the ISO's surface area and we can also take into account that only a fraction $f$ ($f=0\rightarrow1$) of the ISO's sunlit surface may be active. Thus, the sublimation rate of an ISO at $r=1\au$ under these assumptions is given by 
\begin{equation}
\frac{Q({\rm H_2 O})}{\mathrm{molecules\ sec}^{-1}}  
   = 1.1\times10^{28} \, f \, \Biggl[\frac{D}{\km}\Biggr]^2 
   = 4.8\times10^{35} \, f \, 10^{-0.4H}
\end{equation}

The water sublimation rate for new Oort-cloud comets is $\propto  r^{-1.5}$ \citep{Meisel1982} and, assuming this relationship holds within $10\au$ of the Sun, \citet{Jorda2008} found a strong correlation between $Q({\rm H_2O})$ and heliocentric cometary magnitude, $V_C(r)$,  of $\log_{10} \left[Q({\rm H_2 O})\right] = 30.68 - 0.245 \, V_C(r)$ for $r\leq 4.5\au$. The correlation had a scatter of $\sim1\mags$ in $V_C(r)$ which may be due to measurement uncertainties and variations in phase angle scattering for the cometary dust comae that can dominate the apparent magnitude. Inverting the relationship for $V_C(r)$ and substituting the expression for $Q({\rm H_2 O})$ yields
\begin{equation}
V_C(r) = 1.6H - 4.1\log_{10} f + 6.2\log_{10} r - 21.1.
\end{equation}
To provide the most stringent confidence limit on the ISO spatial number density we assume the ISO's entire sunlit hemisphere is active ($f=1$).  Given that $V_C(r)  = V - 5\log_{10} \Delta$, the apparent magnitude of a fresh long-period active comet (an ISO) is 
\begin{equation} 
\label{eq.V_active}
V = 1.6 \, H  + 6.2 \, \log_{10}r + 5 \, \log_{10}\Delta - 21.1
\end{equation}

Our derivation of \eqn{eq.V_active} relies on a chain of assumptions with several caveats.  First, incoming fresh comets and ISOs may be highly active by $r\sim9\au$ \citep{Meech2013}.  This activity must be driven by  volatile species such as CO or CO$_2$ because at this distance H$_2$O is inert, but \eqn{eq.V_active} relies on the activity being water-driven. Second, the rate at which the sublimation rate increases at large $r$ is not well characterized and may vary dramatically between comets, and we have little substantive knowledge of how CO/CO$_2$ sublimation drives the  dust coma and apparent brightness at large heliocentric distances.  Finally, the actual apparent magnitude derived by \citet{Jorda2008} may not correspond to the flux identified by automated detection software that usually detects objects based on the flux within a relatively small aperture with a radius on the order of the system's point spread function (perhaps $1\arcsec$ to $2\arcsec$).  However long-period comets at $r\geq 6\au$ pre-perihelion are generally compact and/or even stellar in nature \citep{Meech2009, Meech2013}, implying little aperture loss  of the comet flux. Thus, to simplify our analysis, we assume that active ISOs turn on at $10\au$ and the measured apparent magnitude follows \eqn{eq.V_active}.  Given that our analysis below provides confidence limits for completely inactive asteroidal ISOs and also 100\% active cometary ISOs, and that there have been no reported detections of any ISO, a more detailed analysis incorporating more complicated cometary behavior is unnecessary.

\section{ISO interstellar spatial number density limit}
\label{s.ISO_spatial_number_density_limit} 

\citet{Jewitt2003} pointed out that ISO number density limits depend on the slope of the ISO size-frequency distribution and the minimum detectable size (maximum detectable absolute magnitude).  Thus, we let the synthetic ISO number density as a function of heliocentric distance (\fig{fig.ISOnumberDensity.vs.heliocentricDistance}), slope of the size-frequency distribution ($\alpha$), and maximum absolute magnitude ($H_{max}$), be expressed as
\begin{equation}
\rho^*(r;\alpha,H_{max}) = f^*(r) \; \rho_{IS}^*(\alpha,H_{max})
\end{equation}
where $\rho^*_{IS}$ is the interstellar number density that we wish to calculate.   Since $N^*(r) = \rho^*(r) \; V^*(r)$ (note that we do not show the dependence of each term on $\alpha$ and $H$ for clarity and remember that we use the asterisk to denote synthetic values) the number of detected synthetic ISOs in the simulation is then
\begin{equation}
N^* = f^*(r) \; \rho^*_{IS} \; V^*.
\label{eq.N_ISO-expected-synthetic}
\end{equation}

Assuming that we can treat the observation statistics as a Poisson distribution, and since the actual number of discovered ISOs is zero, the 90\% confidence limit (CL) on the expected number of ISOs from the model is $N^{CL}=2.3$, \ie\ the expectation value must be 
$\le 2.3$ ISOs or the surveys have a $\ge 90$\% probability of detecting at least one ISO.  Thus, the confidence limit on the interstellar ISO number density ($\rho_{IS}^{CL}$) from the actual surveys is given by
\begin{equation}
N^{CL} = f(r) \; \rho_{IS}^{CL} \; V.
\label{eq.N_ISO-CL}
\end{equation}

Under the assumption that we have developed a reasonable ISO orbit distribution model and survey simulation, $V \approx V^*$ and $\rho_{IS} \approx C \, \rho_{IS}^*$, where $C$ is a normalization constant between the actual and synthetic ISO population, we can use \eqn{eq.N_ISO-expected-synthetic} and \eqn{eq.N_ISO-CL} to solve for
\begin{equation}
  \rho_{IS}^{CL}(\alpha,H_{max}) 
    = \frac{ N^{CL}}{N^*(\alpha,H_{max})} \; \rho_{IS}^*(\alpha,H_{max}).
\label{eq.NCL}
\end{equation}
The denominator is given by \eqn{eqn.N_ISO.vs.alpha+H} and $\rho_{IS}^*(\alpha,H_{max})$ is extracted directly from our synthetic population.

The ISO size-frequency distribution (SFD) is not known but we assume that it can be represented by a function $\propto 10^{\alpha H}$ like most known small body populations.  If the ISOs are generated in a manner similar to the ejection of objects from our solar system during its early formation, then ISOs are ejected from extra-solar systems during their high-mass period when the conditions probably were consistent with the planetesimals being in a self-similar collisional cascade \citep[\eg][]{Dohnanyi1969,OBrien2003}.  Under these conditions the theoretical value of the SFD slope parameter is $\alpha=0.5$.  Deviations from the conditions required for the self-similar collision cascade typically induce `waves' in the SFD \citep[\eg][]{Durda1998,OBrien2003} such that the SFD has a size-dependent slope, but the SFD over a limited range is still exponential.  Thus, we calculated the ISO number density 90\% confidence limits on a grid of $\alpha$ and $H_{max}$ combinations with $0.2\le\alpha\le0.8$ and $10 \le H_{max}\le 20$ in steps of 0.05 in the slope and $1\mags$ in $H$.

For each $(\alpha,H_{max})$ combination we assign the synthetic ISOs random $H$ values distributed according to that SFD and $H_{max}$, compute their new apparent magnitude $V$, then determine the objects' tracklet detection efficiency and digest scores so that we can calculate the total number of objects that would have been detected with $H<H_{max}$ if the SFD has slope $\alpha$.  Since each synthetic ISO was randomly assigned a different $H$ value we repeated the procedure $10\times$ for each slope parameter and averaged the results.  Changing the absolute magnitude for each object has the effect of changing its apparent magnitude which directly affects the efficiency for detecting the object and alters its digest score.  The $10\times$ repetition was determined empirically to reduce the statistical noise in $\rho_{IS}^{CL}(\alpha,H_{max})$.

\section{Results \& Discussion} 
\label{s.Discussion} 

\begin{figure}[htbp]
\begin{center}
  \includegraphics[width=.45\textwidth]{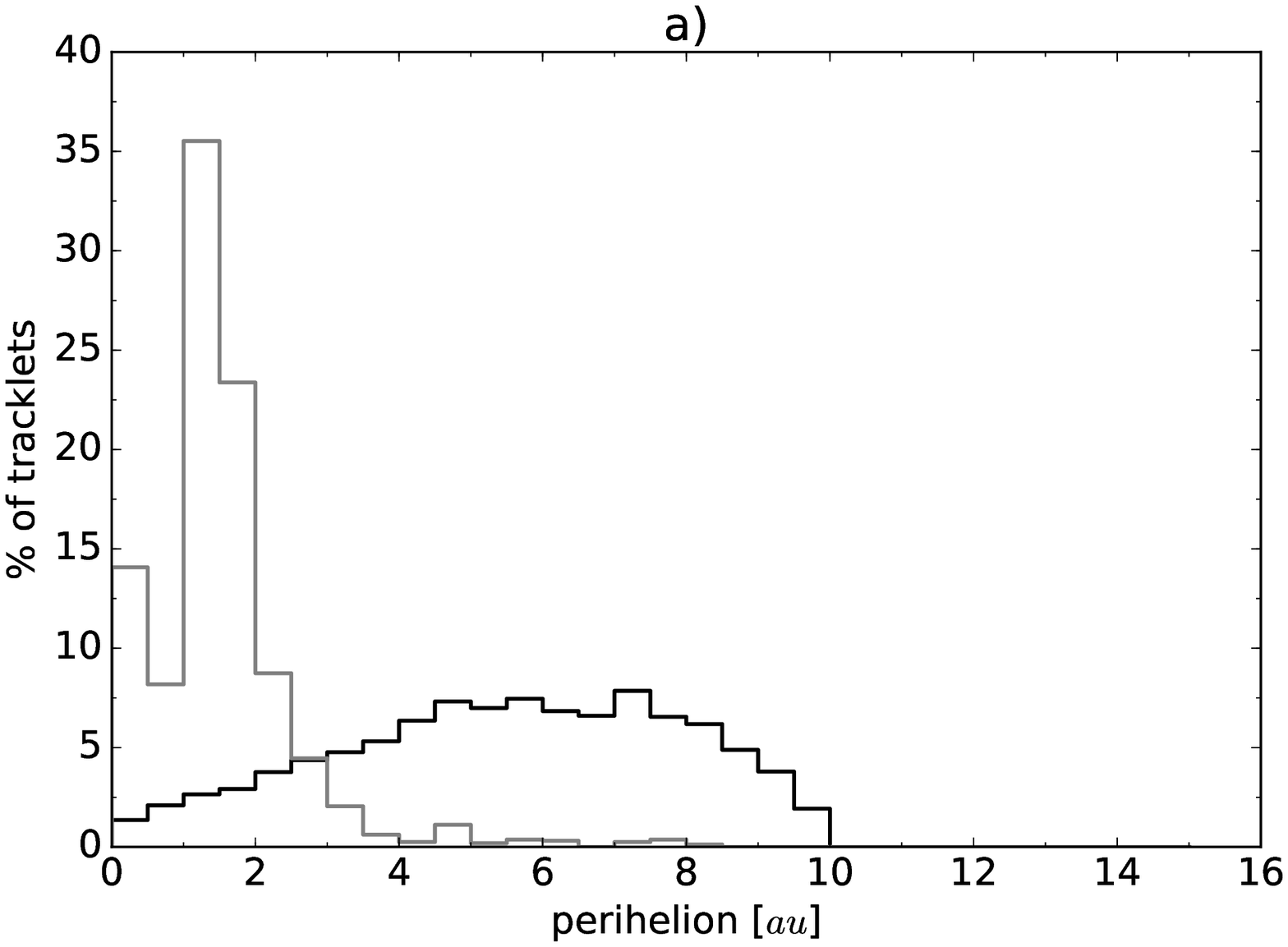}
  \includegraphics[width=.45\textwidth]{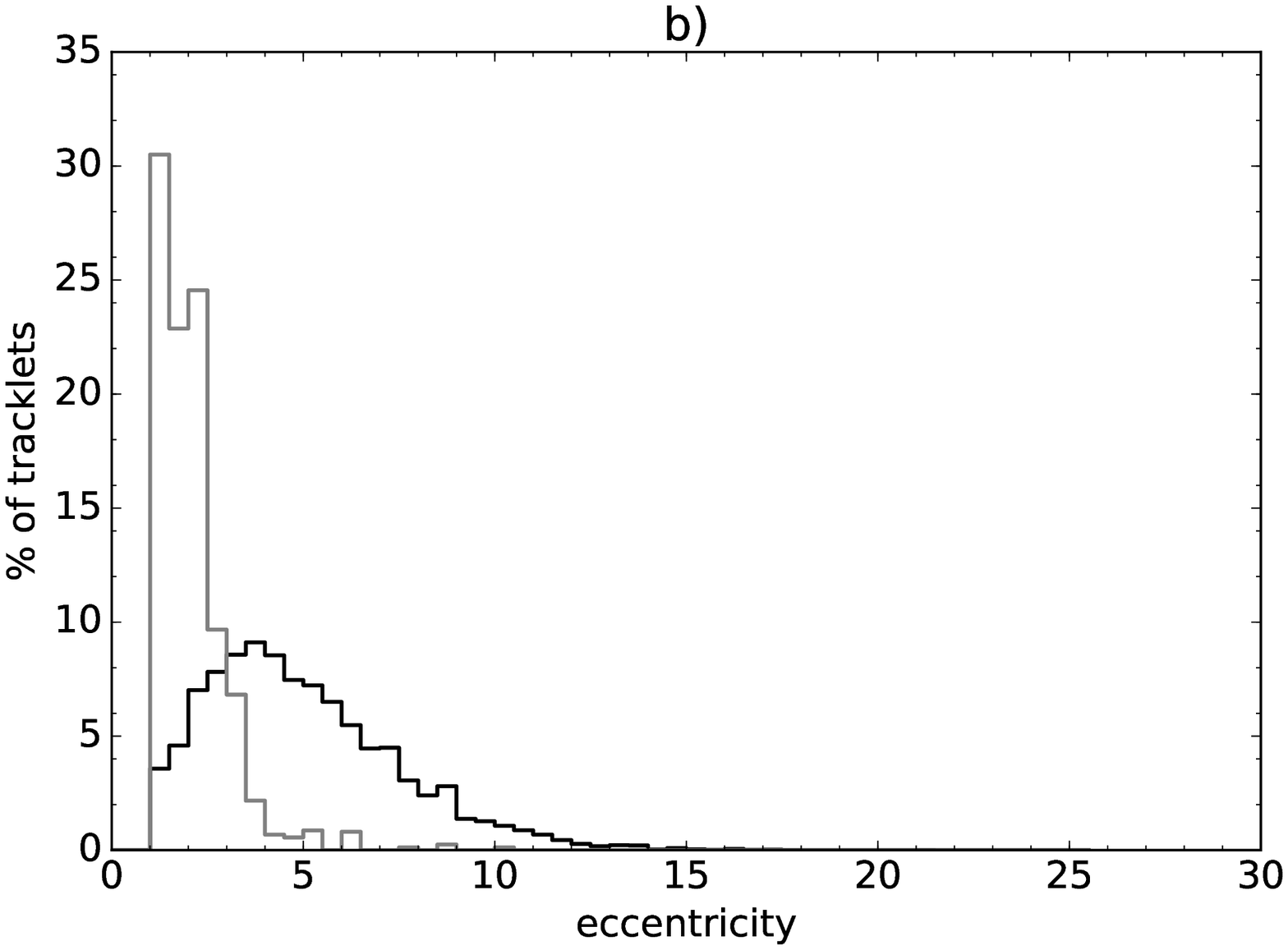}
  \includegraphics[width=.45\textwidth]{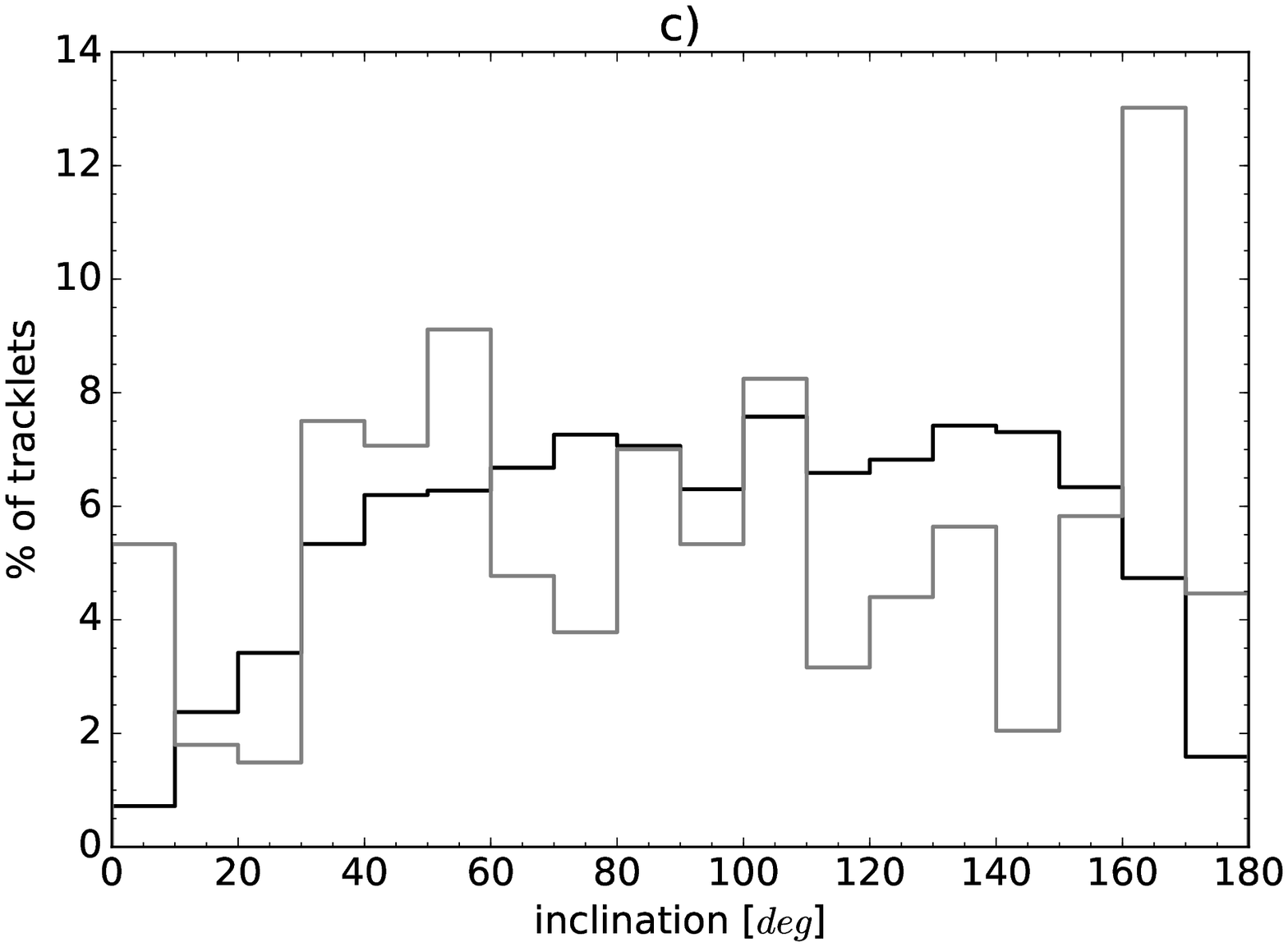}
\caption{Orbit element distributions for synthetic detected ISOs for a \PSone\ simulation at the nominal values of $\alpha=0.5$ and $H_{max}=19$ as described in the text and indicated in \fig{fig.ISO-NumberDensityCL-measured-vs-alpha-Hmax}.  The black lines represent the distributions for `active' comets (\eqn{eq.V_active}) and the gray lines represent the case of inactive asteroids.  We require that the `digest' score be $>90$ in both cases.}
\label{fig.PS1-detected-qei}
\end{center}
\end{figure}

The synthetic detected ISOs have very different orbit element distributions from the generated population due to observational selection effects (\fig{fig.PS1-detected-qei}).  

The perihelion distribution (\fig{fig.PS1-detected-qei}a) is very different between inactive and active ISOs with the asteroidal ISOs typically being detected at about the distance of Mars while the cometary ISOs are typically detected between the distance of Jupiter and Saturn with an as-designed cutoff at $10\au$, the distance at which we assume the onset of cometary activity.  This behavior is explicable because with $H_{max}=19$ and $\alpha=0.5$ most of the detected objects will be in the $\km$ diameter range and must be within about $2\au$ to be detectable by the modeled surveys.  Since the comets are active they are much brighter at the same absolute magnitude and can therefore be detected at much larger distances.  

The eccentricity distribution of the detected objects (\fig{fig.PS1-detected-qei}b) is strongly skewed to small eccentricities due to the previously mentioned correlation between an ISO's perihelion distance and eccentricity.  Since the asteroidal ISOs have smaller perihelion distances their eccentricities are more skewed towards $e \ga 1$ than the cometary ISOs.  Whereas the most probable eccentricity in the generated ISO population is $\sim 270$ the detected synthetic ISOs have $\sim100\times$ smaller modes of $\sim 1.4$ and $\sim 3.5$ for inactive and active ISOs respectively. 

The smallest eccentricity in the detected synthetic ISO cometary population has $e \sim 1.01$, smaller than comet C/1980~E1~(Bowell) that has an eccentricity\footnote{according to the JPL Small-Body Database Search Engine as of 2016 December 13} of $\sim 1.0577$.  Indeed, five known comets have $e \ge 1.01$, suggesting that, based only on their orbital eccentricity, it is possible, but not at all likely as described above, that these objects could be ISOs because our work shows that ISOs that have small perihelion distances will also have small eccentricities and may appear to be slightly perturbed Oort cloud comets.

Finally, the inclination distribution of the synthetic detected objects (\fig{fig.PS1-detected-qei}c) retains the general shape of the $\sin i$ distribution of the underlying generated population for both the asteroidal and cometary ISOs.  The distributions are slightly skewed to retrograde orbits by the requirement that the digest score be $>90$ to flag the object as interesting enough to trigger a followup campaign and the retrograde orbits are `easier' to flag as unusual.

\begin{figure}[!ht]
  \centering	
  \includegraphics[width=.45\textwidth]{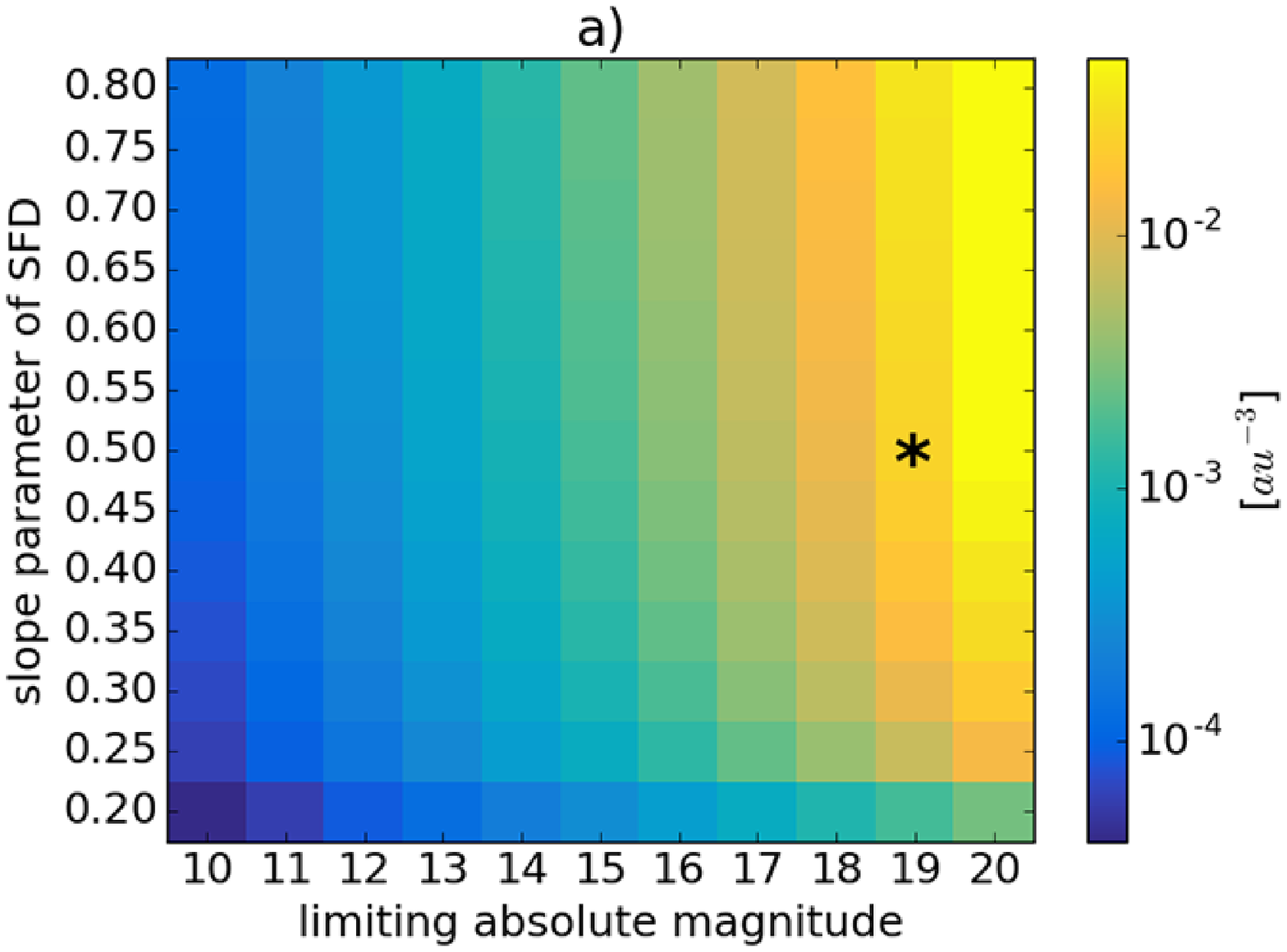}
  \includegraphics[width=.45\textwidth]{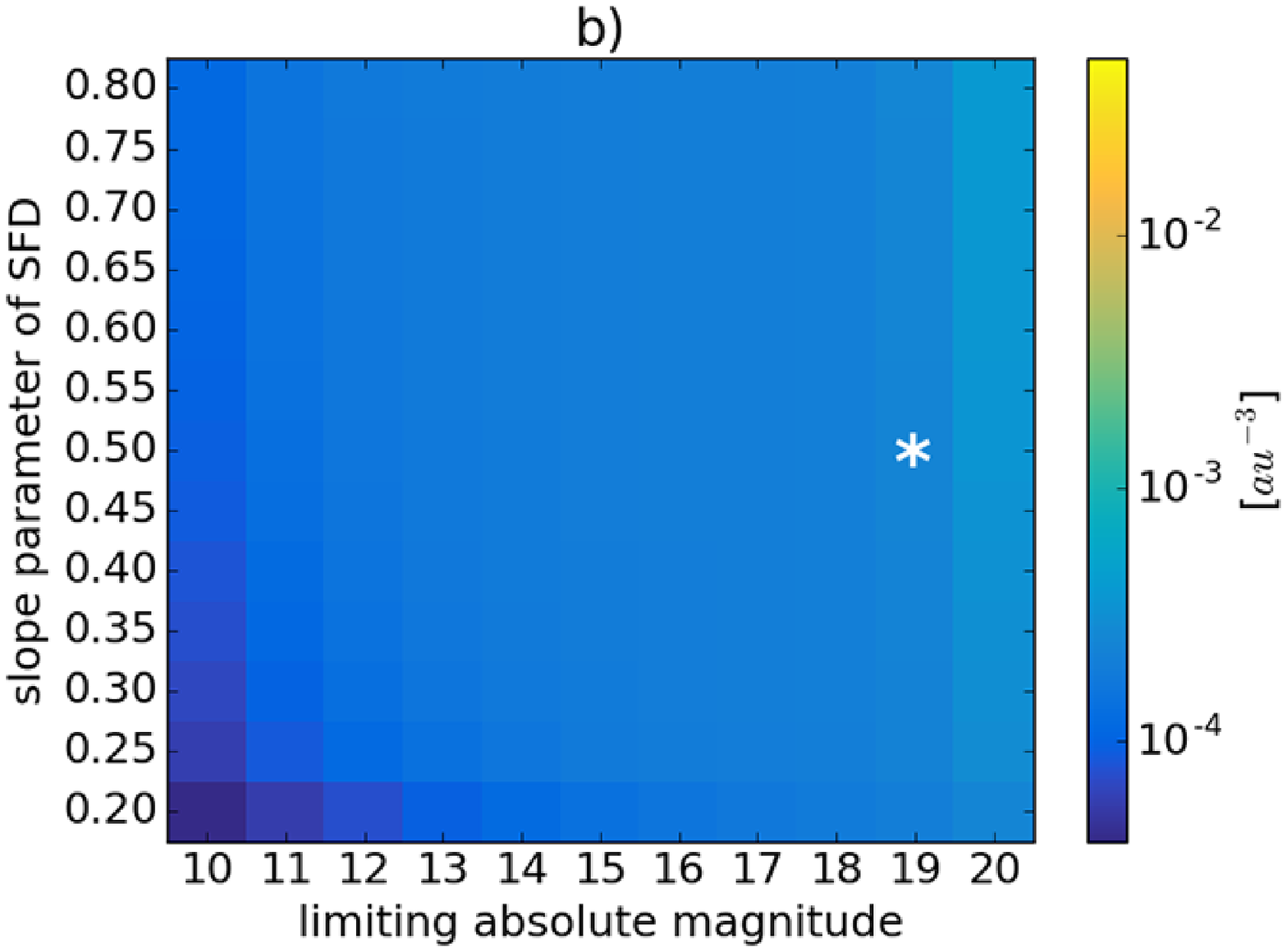}
  \caption{90\% confidence limit on the ISO number density versus SFD slope parameter $\alpha$ and limiting absolute magnitude $H_{max}$ ({\bf left}) without cometary activity and ({\bf right}) with cometary activity.  The asterisk at a slope parameter of $\alpha=0.5$ and limiting absolute magnitude $H=19$ correspond to the canonical slope for self-similar cascade \citep{Dohnanyi1969} and a $1\km$ diameter ($H=19.1$) comet with an albedo of $p_V=0.04$ (\eqn{eq.Diameter.vs.albedo+H}).}
  \label{fig.ISO-NumberDensityCL-measured-vs-alpha-Hmax}			
\end{figure}

Our 90\% confidence limit on the interstellar ISO number density improves, \ie\ is numerically smaller, as $H_{max}$ and $\alpha$ decrease (\fig{fig.ISO-NumberDensityCL-measured-vs-alpha-Hmax}a).  This is because the distance at which an ISO is detectable increases as the maximum detectable ISO diameter decreases ($H_{max}$ increases) but not fast enough to compensate for the slope of the SFD and the apparent brightness decreasing like $\Delta^4$.  The CL improves with a shallower SFD slope because a larger fraction of the ISOs are large and bright enough to be detected.  Furthermore, the CL improves dramatically if we assume that the ISOs display cometary activity as they will be much brighter, and therefore more easily detected, at heliocentric distances up to $10\au$, the distance at which comets become active in our model  (\fig{fig.ISO-NumberDensityCL-measured-vs-alpha-Hmax}b).  The CLs with and without cometary activity represent the full range of CLs in our analysis from about $10^{-4}\au^{-3}$ to $10^{-1}\au^{-3}$ over the $(\alpha,H_{max})$ range.  

\begin{figure}[!hbt]
  \centering
  \includegraphics[width=0.95\textwidth]{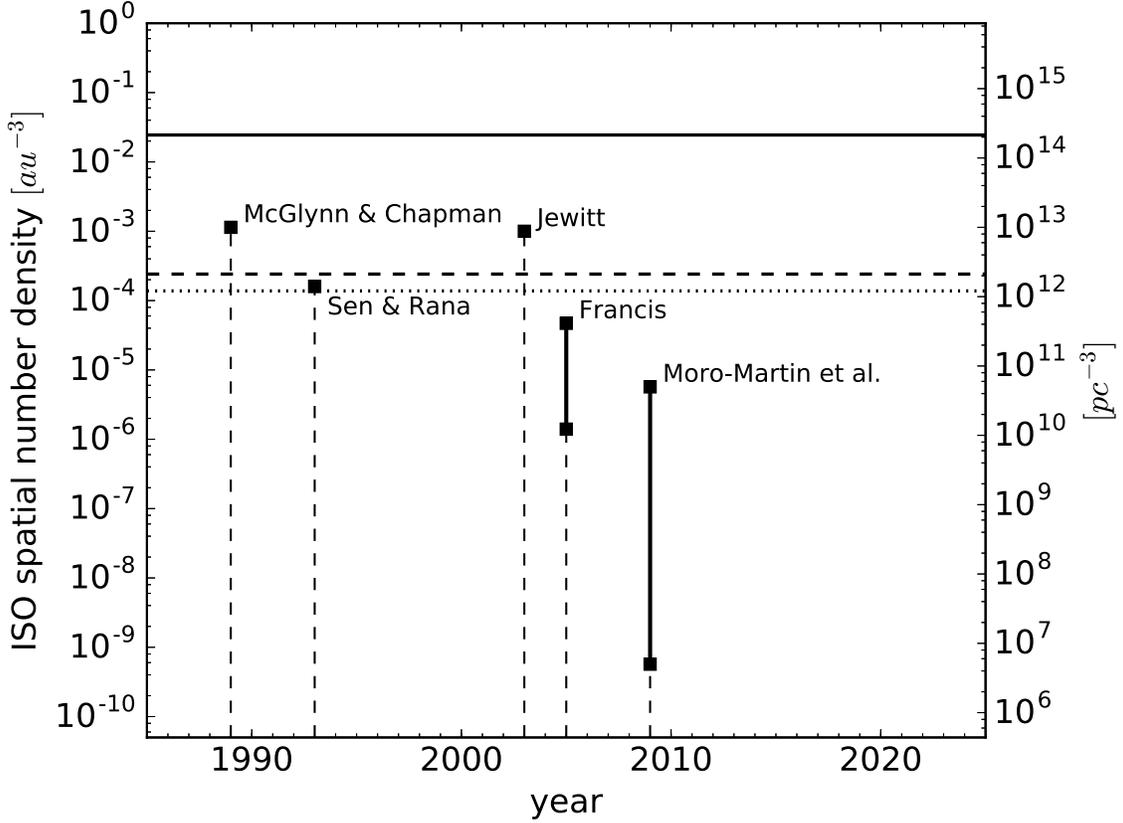}
  \caption{Our ISO interstellar number density 90\% confidence limit for three combined surveys and other theoretical values as a function of time.  The uppermost solid line represents the confidence limit assuming that the ISOs do not show any cometary activity ($2.4\times10^{-2}\au^{-3}$) while the dashed line represents the limit assuming that 100\% of the ISO's sun-facing surface is active ($2.4\times10^{-4}\au^{-3}$). The dotted line illustrates the limit assuming that all ISO candidates would be identified even though their NEO digest score may not exceed the NEO threshold \ie\ the ISO candidates could be identified by their non-stellar point-spread function rather than their unusual rates of motion ($1.4\times10^{-4}\au^{-3}$). 
}
  \label{fig.NumberDensityCL-vs-predictions}
\end{figure}

To compare our CLs with theoretical predictions for the interstellar ISO number density we use the CLs corresponding to canonical values of $H_{max}=19.1$ (about $1\km$ diameter) and the \citet{Dohnanyi1969} SFD slope of $\alpha=0.5$.  Objects of this absolute magnitude are detectable near opposition at heliocentric distances of about $2.2\au$, $2.1\au$, and $1.7\au$ with the \PSone\  (\wps\ filter), MLS and CSS surveys respectively (using the limiting magnitudes provided in \tab{tab.SurveyEfficiencyParameters}).  The 90\% CL for asteroidal photometric behavior at the canonical values is $2.4\times10^{-2}\au^{-3}$.  Given that solar gravitational focussing yields ISO spatial number densities about $2\times$ the interstellar value in the range $1\au\la r \la 3\au$ (\fig{fig.ISOnumberDensity.vs.heliocentricDistance}), at 90\% CL there must be $\la 5$ inactive ISOs within $3\au$ of the Sun at any time or one of the three surveys would have identified an ISO.  The CL at our nominal $(\alpha,H_{max})$ values improves by 2 orders of magnitude to $2.4 \times 10^{-4}\au^{-3}$ if we assume that the ISOs' Sun-facing surfaces are 100\% active and show cometary activity that make them significantly brighter than asteroids of the same size in the same geometrical configuration.  In this case, there must be $\la 0.05$ active ISOs within $3\au$ of the Sun at any time or one of the 3 surveys would have identified an ISO.

The three asteroid surveys working in tandem over a cumulative 19 years can not set an interesting limit on the interstellar ISO spatial number density if the ISOs behave more like asteroids than comets (\fig{fig.NumberDensityCL-vs-predictions}).  In this case, the limit is an order of magnitude higher than even the most optimistic predictions.  However, we consider it unlikely that the ISOs will act more like asteroids than comets given that only a small percentage of objects with long period orbits in our solar system are inactive (since 2006 there have been 107 LPC discoveries of which only 2 have no measurable activity\footnote{\,Meech, K.~J., University of Hawai'i, personal communication}). \PSone\ is currently the leading discoverer system of these nearly inactive `Manx' objects \citep{Meech2016} so we expect it would efficiently discover even inactive ISOs passing through our solar system.

Under the assumption of cometary activity in ISOs our 90\% CL on the maximum interstellar ISO spatial number density of $2.4 \times 10^{-4}\au^{-3}$ is considerably lower than the predictions of \citet{Jewitt2003} and \citet{McGlynn1989} but still higher than the upper limit determined by \citet{Francis2005}.  Thus, this CL and \citet{Francis2005} suggest that the asteroid surveys are beginning to probe an ISO spatial number density range that could have implications for planetary formation.  Taken at face value, and in comparison to the predicted values, the two CLs imply that other solar systems are not similar to our own in terms of the ejection of proto-planetary material, that the ISOs have an unexpected SFD, or perhaps that the ISOs are not distributed homogeneously throughout the galaxy.  

The agreement between our CL {\it with} cometary activity and that of \citet{Francis2005} is surprising given that he also assumed that the LINEAR survey would identify comets by their unusual rates of motion (or they would not be reported as NEO candidates for followup) and would detect them at `cometary' distances due to their increased brightness.  It is surprising because our analysis uses the results from 3 surveys, 2 of which have much deeper limiting magnitudes than LINEAR, over 19\,years compared to the LINEAR data of 3\,years.  We attribute the difference in the CLs to our use of a more sophisticated ISO model and our direct access to the pointing history and detection efficiency estimates for our three surveys.   Also, \citet{Francis2005} suggested that the ``results are quite sensitive to the adopted bright-end slope of the absolute magnitude distribution''.  While our analysis explored the entire range of SFD slopes, we quote the CL for $\alpha=0.5$ and with $H_{max}=19.1$, whereas \citet{Francis2005} used his own long-period comet SFD for the ISOs and suggests that most of the detected ISOs will have $H\sim6.5$ (almost $200\km$ diameter).

Our most stringent $\rho_{IS}^{CL} = 1.4 \times 10^{-4}\au^{-3}$ when $\alpha=0.5$ and $H_{max}=19.1$ assumes that all ISOs with an apparent magnitude brighter than the survey system's limiting magnitude will be identified as candidate ISOs by their morphology instead of through their unusual rates of motion at their sky-plane location (\ie\ their NEO digest scores).  This ISO identification scenario is reasonable because 1) ISOs are likely to display cometary behavior as described above and 2) the \PSone\ manual vetting of each detection ensures that each tracklet has been reviewed by an observer trained at discerning even weak cometary behavior \citep{Hsieh2015}.\footnote{Manual \PSone\ vetting of known comets indicates that the realized comet detection efficiency was $\la70$\%  \citep{Hsieh2015}.} 
The assumption that the ISOs will display cometary activity and be detected by the survey system results in a 90\% CL that is more than 2 orders of magnitude lower than the limit assuming no activity.  Furthermore, a $1\km$ diameter ($H_{max}=19.1$) comet has an effective cometary absolute magnitude corresponding to a much larger object that should render them detectable at heliocentric distances where cometary activity turns on due to volatile sublimation, about $10\au$.  The $\rho_{IS}^{CL} = 1.4 \times 10^{-4}\au^{-3}$ value suggests that there are, very roughly, less than about 0.5 cometary ISOs within about $10\au$ of the Sun, \ie\ within a heliocentric sphere with a radius comparable to Saturn's distance from the Sun.  This CL is on the threshold of being able to reject (\fig{fig.NumberDensityCL-vs-predictions}) the ISO spatial density prediction of \citet{Sen1993}.

Our interstellar ISO limit is almost an order of magnitude smaller than the $\rho_{IS}=10^{-3}\au^{-3}$ value used by \citet{Torbett1986} to predict that the ISO capture rate by Jupiter is about once per 60~Myr.  Assuming that the capture rate scales with the ISO spatial density, our CL suggests that the ISO capture rate into heliocentric orbit is less than about once per $\sim 400$\,Myr, making it less likely that the unusual comet 96P/Machholz is an interstellar interloper.

\citet{Moro-Martin2009}'s theoretical interstellar ISO number density prediction included several enhancements beyond the earlier estimates with the most important being 1) a stellar-mass-dependent stellar number density, 2) stellar-mass-dependent protoplanetary disk mass, 3) the fraction of stars that harbor the giant planets necessary to scatter planetesimals into interstellar space, and 4) the ISO size-frequency distribution.  Their detailed analysis dramatically reduces the expected interstellar ISO number density to the range from about $10^{-6}\au^{-3}$ to $10^{-10}\au^{-3}$, many orders of magnitude smaller than our best experimental limit.  If the actual ISO spatial number density lies somewhere in that range it will be essentially impossible for the three surveys used in this analysis to ever detect ISOs barring a statistical fluke.  Thus, the detection of the first non-microscopic ISOs will require new survey systems like the LSST \citep[\eg][]{Ivezic2008} --- and some luck. Despite LSST's nominal 10-year mission that will deliver about $320\meter^2$-years of surveying (the product of its effective aperture area and the survey time), $\sim13\times$ more than the three combined surveys in this analysis, \citet{Cook2011} suggest that LSST will not detect any ISOs beyond $5\au$ and the expected number within that distance is small.

\section{Conclusions}
\label{s.Conclusions}

The prospects for identifying a large chunk of material ejected by an extra-solar system passing through our own solar system appear to be bleak.  The fact that the existing multi-year asteroid surveys, \PSone, Catalina Sky Survey, and the Mt. Lemmon Survey, have not yet identified an ISO indicates either that other solar systems do not form like ours, ejecting the vast majority of resident material in the process, or that the ISO size distribution does not approximate that expected for a self-similar collisional cascade as (roughly) observed for populations of small bodies in our solar system.  Our most stringent 90\% upper confidence limit on the interstellar number density of interstellar objects larger than $1\km$ diameter is $1.4\times10^{-4}\au^{-3}$ which assumes that the ISOs will display a behavior typical of first time active comets entering the solar system with outgassing and associated increased apparent brightness beginning at about $10\au$ from the Sun.  In this case the object may be detected as cometary through its morphological appearance in the image even though its digest score may not be interesting.

ISOs can have very small eccentricities approaching $e=1$ for parabolic orbits, especially for objects with small perihelion that are more efficiently detected by astronomical surveys. Roughly 0.003\% of the model ISOs in our simulation had $1.01 \le e \le 1.06$, a range in which five comets are also known. While it is more likely that the $e>1$ values for the known objects are due to planetary perturbations or astrometric errors our results suggest that care should be taken before automatically rejecting $e>1$ objects as ISO candidates.

Finally, our results suggest that if an ISO passed through our solar system and was detectable by contemporary surveys that there is still a $\sim 35$\% probability that it was not identified as an ISO due to lack of followup. Future sky surveys like the LSST will have higher ISO detection efficiency due to their regular self-followup surveying and automated tracklet linking and orbit determination.

\section*{Acknowledgements}

We thank Henry Hsieh for helping us understand \PSone\ comet detection efficiency and Urs Hugentobler for reviewing and providing many suggestions to guide the work.  Peter Brown was helpful in understanding interstellar meteor data.  Dan Tamayo was very helpful in providing information and updates to their REBOUND software \citep{Rein2012} to handle hyperbolic orbits.  Robert Weryk kindly provided assistance in determining the number of \PSone\ objects that never have followup observations.  An anonymous reviewer provided helpful suggestions to improve the manuscript.  We thank the PS1 Builders and PS1 operations staff for construction and operation of the PS1 system.

Peter Vere\v{s}'s Pan-STARRS MOPS Postdoctoral Fellowship at the University of Hawai`i's Institute for Astronomy was sponsored by NASA NEOO grant No. NNX12AR65G.  Some of this research was conducted while he was employed at the Jet Propulsion Laboratory, California Institute of Technology, under a contract with the National Aeronautics and Space Administration.  All rights reserved.  Alan Fitzsimmons acknowledges support from STFC grant ST/L000709/1. The CSS is currently supported by NASA Near Earth Object Observations program grant NNX15AF79G, "The Catalina Sky Survey for Near Earth Objects".  The Pan-STARRS1 Surveys (PS1) have been made possible through contributions of the Institute for Astronomy, the University of Hawaii, the Pan-STARRS Project Office, the Max-Planck Society and its participating institutes, the Max Planck Institute for Astronomy, Heidelberg and the Max Planck Institute for Extraterrestrial Physics, Garching, The Johns Hopkins University, Durham University, the University of Edinburgh, Queen's University Belfast, the Harvard-Smithsonian Center for Astrophysics, the Las Cumbres Observatory Global Telescope Network Incorporated, the National Central University of Taiwan, the Space Telescope Science Institute, the National Aeronautics and Space Administration under Grant No. NNX08AR22G issued through the Planetary Science Division of the NASA Science Mission Directorate, the National Science Foundation under Grant No. AST-1238877, the University of Maryland, and Eotvos Lorand University (ELTE) and the Los Alamos National Laboratory.

\bibliography{references.bib}
\bibliographystyle{icarus}

\end{document}